\documentclass[aps,prd,twocolumn,showpacs,superscriptaddress,letterpaper,asmmath,amssymb]{revtex4}

\usepackage{graphicx}% Include figure files
\usepackage{dcolumn}% Align table columns on decimal point
\usepackage{bm}% bold math
\usepackage{feynmf}%%%{feynmp}

\RequirePackage{xspace}
\usepackage{relsize}

\def\W      {\ensuremath{W}\xspace}

\def\t     {\ensuremath{t}\xspace}

\def\Kbar  {\kern 0.2em\overline{\kern -0.2em K}{}\xspace}

\def\Bbar    {\kern 0.18em\overline{\kern -0.18em B}{}\xspace}

\def\tprime  {\ensuremath{t^{\prime}}\xspace}
\def\bprime  {\ensuremath{b^{\prime}}\xspace}

\def\missET {{\not\!\! E_T}}
\def\Qbar    {\kern 0.08em\overline{\kern -0.08em Q}{}\xspace}

\def\lfj {\ensuremath{\ell+4j}\xspace}
\def\lfivej {\ensuremath{\ell+5j}\xspace}
\def\lljbme {\ensuremath{\ell^{\pm}\ell^{\pm}jb\missET}\xspace}

\def\invfb   {\ensuremath{\mbox{\,fb}^{-1}}\xspace}
\newcommand{\mev}{\ensuremath{\mathrm{\,Me\kern -0.1em V}}\xspace}
\newcommand{\mevc}{\ensuremath{{\mathrm{\,Me\kern -0.1em V\!/}c}}\xspace}
\newcommand{\mevcc}{\ensuremath{{\mathrm{\,Me\kern -0.1em V\!/}c^2}}\xspace}
\newcommand{\gev}{\ensuremath{\mathrm{\,Ge\kern -0.1em V}}\xspace}
\newcommand{\gevc}{\ensuremath{{\mathrm{\,Ge\kern -0.1em V\!/}c}}\xspace}
\newcommand{\gevcnospace}{\ensuremath{{\mathrm{\,Ge\kern -0.1em V\!/}c}}}
\newcommand{\gevcc}{\ensuremath{{\mathrm{\,Ge\kern -0.1em V\!/}c^2}}\xspace}
\newcommand{\bea}{\begin{eqnarray}}
\newcommand{\eea}{\end{eqnarray}}

\def\missET {{\not\!\! E_T}}

\begin{document}
%\pagewiselinenumbers
% the following line is for submission
%\hspace{5.2in} \mbox{Fermilab-Pub-04/xxx-E}
\bibliographystyle{apsrev}

\title{Fourth generation quark mass limits in CKM-element space}
%\title{{Search for fermion-pair decays}}
%\\ \vspace*{2.0cm}}

\author{Christian J. Flacco}
\email{cflacco@mac.com}
\affiliation{Department of Physics and Astronomy, University of California, %
Irvine, CA 92697, USA}
\author{Daniel Whiteson}
\email{daniel@uci.edu}
\affiliation{Department of Physics and Astronomy, University of California, %
Irvine, CA 92697, USA}
\author{Matthew Kelly}
\email{mkelly@uci.edu}
\affiliation{Department of Physics and Astronomy, University of California, %
Irvine, CA 92697, USA}

\begin{abstract}
We present a reanalysis of CDF data to extend limits on individual
fourth-generation quark masses from particular flavor-mixing rates to
the entire space of possible mixing values.  Measurements from CDF have set individual limits on 
masses, $m_\bprime$ and $m_\tprime$, at the level of $335$--$385$ GeV
assuming specific and favorable flavor-mixing rates. We consider the
space of possible values for the mixing rates and find that the CDF data imply limits of $290$ GeV and greater over a wide range of mixing
scenarios. We also analyze the limits from the perspective of a four-generation CKM matrix. We find that 
present experimental constraints on CKM elements do not suggest further constraints on fourth-generation
quark masses.
\end{abstract}

% 12.60.-i Models beyond the standard model
% 13.85.Rm Limits on production of particles
% 14.80.-j Other particles (including hypothetical)


\maketitle

\date{\today}

\section{Introduction}

A simple modification of the standard model is the
addition of a fourth sequential generation of fermion doublets.  This
natural extension of the standard model~\cite{reviews} may trigger dynamical
electroweak symmetry breaking~\cite{DEWSB} without a Higgs boson, and so address the
hierarchy problem.  The new heavy fermions may have Yukawa couplings
so large that they become strong. The subsequent strong dynamics may lead to a
composite of fourth generation fermions performing the role of the Higgs~\cite{hung,hashimoto1,hashimoto2}.

Recent searches by the CDF Collaboration for direct production of the fourth generation quarks, denoted $t'$ and $b'$
for the up- and down-type,
found $m_{t'}>335$ GeV\cite{CDFt} and $m_{b'}>385$ GeV \cite{CDFb2}, assuming 
$\mathcal{B}(t'\rightarrow W\{q=d,s,b\})=100$\% and
$\mathcal{B}(b'\rightarrow Wt)=100$\% respectively. This suggests that fourth generation fermions must indeed
be heavy, in support of the compositeness scenario.
These searches
typically have been interpreted under the assumptions of 
$m_{t'} - m_{b'} < M_W$ and negligible mixing of
the $(t',b')$ states with the two lightest quark generations. 
Such conditions are generally required for the SM4 with one Higgs 
doublet, to account for EW precision data \cite{EWPD}.

Moreover, when a fourth generation of fermions is embedded
in theories beyond the SM, the large splitting case ($m_{t'} - m_{b'} > M_W$) and the inverted
scenario ($m_{t'} < m_{b'}$) have not been excluded. An example of this was given recently
\cite{hashimoto2}, showing that precision EW data can accomodate $m_{t'} - m_{b'} > M_W$ 
if there are two Higgs doublets. 
In fact, the
compositeness picture emerging from the addition
of new heavy fermionic degrees of freedom is more
naturally described at low energies by multi-Higgs
theories \cite{hung,hashimoto1,hashimoto2}.

This subtle theoretical landscape suggests that there is no 
uniquely interesting set of assumptions under which
experimental data must be interpreted. From the experimental
view point, choosing a simple set of assumptions allows straightforward,
if narrow, interpretation of results. Yet these interpretations may be
extended\cite{lim_prl} to give broader and more general results.
This article continues the work of our previous Letter~\cite{lim_prl};
we again consider the possible 4th generation flavor-mixing space
broadly, and apply results of the previous searches by CDF to calculate direct
limits for arbitrary mixing values and for both cases of $|m_{t'} -
m_{b'}| > M_W$.  This article includes significant unpublished details of the
previous calculations, an additional recently published third dataset~\cite{CDFb2}, and a translation of the mass limits that we derive in
branching fraction space into mass limits in flavor mixing space, which allows for direct comparison
with other experimental flavor mixing constraints.

The following sections describe the original CDF measurements, mass
limits in branching fraction space, and the limits remapped to CKM4 space.

\section{Samples and Strategy}

The CDF Collaboration has published several important limits on fourth-generation quark masses.
We orient this discussion with a summary of their samples and results.  Three analyses have been presented:  
(1) a collection of events containing at least four jets
and a single lepton, known as the \lfj sample\cite{CDFt}; and (2) a
collection of events containing a single lepton, and at least five
jets (one with a flavor tag), known as the \lfivej sample\cite{CDFb2}; and (3) a collection of events 
containing two same-charge leptons, two jets (one with a flavor tag) and 
evidence of neutrinos (missing 
transverse energy), known as the  \lljbme sample \cite{CDFb}.

\subsection{The \lfj sample}
In 4.6\invfb of data, CDF searched for $t'\rightarrow W\{q=d,s,b\}$ decays in the mode 
%(shown in Fig.~\ref{fig:wq_diag})
\[ p \bar p \rightarrow  t'\bar{t'}\rightarrow (W\rightarrow l\nu) q(W\rightarrow qq')q \]
by requiring a single lepton and at
least four jets. The data were analyzed by reconstructing the invariant
mass of the candidate $t'$ and measuring the total energy in
the event. The event selection used the four jets of highest transverse energy in the event, but
did not require a flavor signature ($b$-tag) on any of the jets, making it generally
sensitive to $t'\rightarrow W\{q=d,s,b\}$.  Assuming $\mathcal{B}(t'\rightarrow
W\{q=d,s,b\})=100\%$, CDF found $m_t' > 335$ GeV.

The mass reconstruction used minimum-likelihood fitting methods that depend upon the particular
spectra of final state components.  If, for
example, one half of an event decayed as $t'\rightarrow W(b'
\rightarrow Wq)$ giving a $WWqWq$ topology (rather than $t'\rightarrow
Wq$ giving the $WqWq$ topology), it might satisfy
 the $\ell+4j$ selection criteria, but the reconstructed mass
distribution for such events would be significantly modified by the additional
$W$. Thus, the results cannot
be trivially applied to topologies other than $WqWq$, even when they are expected
contributors to the studied sample. We therefore apply the $\ell+4j$ results to $WqWq$ 
processes {\it exclusively}.

\subsection{The \lfivej sample}

In 4.8fb$^{-1}$ of data, CDF searched for $b'\rightarrow Wt$ decays in
the mode

\begin{eqnarray}
p \bar p \rightarrow b'\bar{b'} \rightarrow WtW\bar{t} & \rightarrow & WWbWW\bar{b} \nonumber \\
 & \rightarrow & ( \ell^{\pm} \nu)(qq')b(qq')(qq')\bar{b} \nonumber
\end{eqnarray}
by requiring at least one lepton, at least five jets (one with a $b$
flavor tag), and missing transverse energy of at least 20 GeV.  The
data were analyzed by examining the number of jets in the event and
the total scalar transverse energy in the event ($H_T$).  
Assuming
$\mathcal{B}(b'\rightarrow Wt)=100\%$, CDF found $m_b' > 385$ GeV
with this signature, the highest limit to date.

As in the
\lfj sample, the analysis uses minimum-likelihood fitting methods
which requires knowledge of the distribution of the signal and
background in the $H_T$ variable as well as jet multiplicity.
Reinterpretting these results to set limits on mixtures of
$b'\rightarrow Wt$ and $t'\rightarrow Wb'$ is not possible without
knowledge of these distrubutions, so we apply the \lfivej results
{\it exclusively} to the $WtWt$ processes.

\subsection{The \lljbme Sample}
CDF also searched for $b'\rightarrow Wt$ decays in 2.7\invfb of data, in the
same-charge lepton mode%(shown in Fig.~\ref{fig:wq_diag} ):
\begin{eqnarray}
p \bar p \rightarrow b'\bar{b'} \rightarrow WtW\bar{t} & \rightarrow & WWbWW\bar{b} \nonumber \\
 & \rightarrow & ( \ell^{\pm} \nu)(qq')b(qq')(\ell^{\pm}
\nu)\bar{b} \nonumber
\end{eqnarray}
by requiring two same-charge
leptons, at least two jets (at least one with a $b$-tag), and missing
transverse energy of at least 20 GeV.

 Given the small backgrounds, multiple neutrinos and large jet multiplicity in the sample, CDF did not
reconstruct the $b'$ mass, but instead fit the observed jet
multiplicity to signal and background templates generated from simulations. Assuming
$\mathcal{B}(b'\rightarrow Wt)=100\%$, CDF found $m_b' > 338$ GeV.

Since the $\ell^{\pm}\ell^{\pm}jb\missET$ analysis did not use final-state dependent
fits, these results are process independent---they may be applied to any process producing 
the $\ell^{\pm}\ell^{\pm}jb\missET$ signal. For example, $\tprime\rightarrow\W\bprime\rightarrow
\W\W\t\rightarrow\W\W\W b$ decays would produce a six-$W$, two-$b$ signature,
with higher jet multiplicity and
larger acceptance to the $\ell^{\pm}\ell^{\pm}jb\missET$ sample than
the simple four-$W$, two-$b$ signature. In this analysis, we therefore apply
the $\ell^{\pm}\ell^{\pm}jb\missET$ results to four-$W$, two-$b$
processes {\it inclusively}.

\subsection{Analytic Extension of the CDF Results}

The three data samples can be seen as probing a region of a two-dimensional interval
in branching fraction space which, for the classical splitting ($t'$ heavier than $b'$), is specified in the following way.

\begin{figure}[ht]
\includegraphics[width=0.7\linewidth]{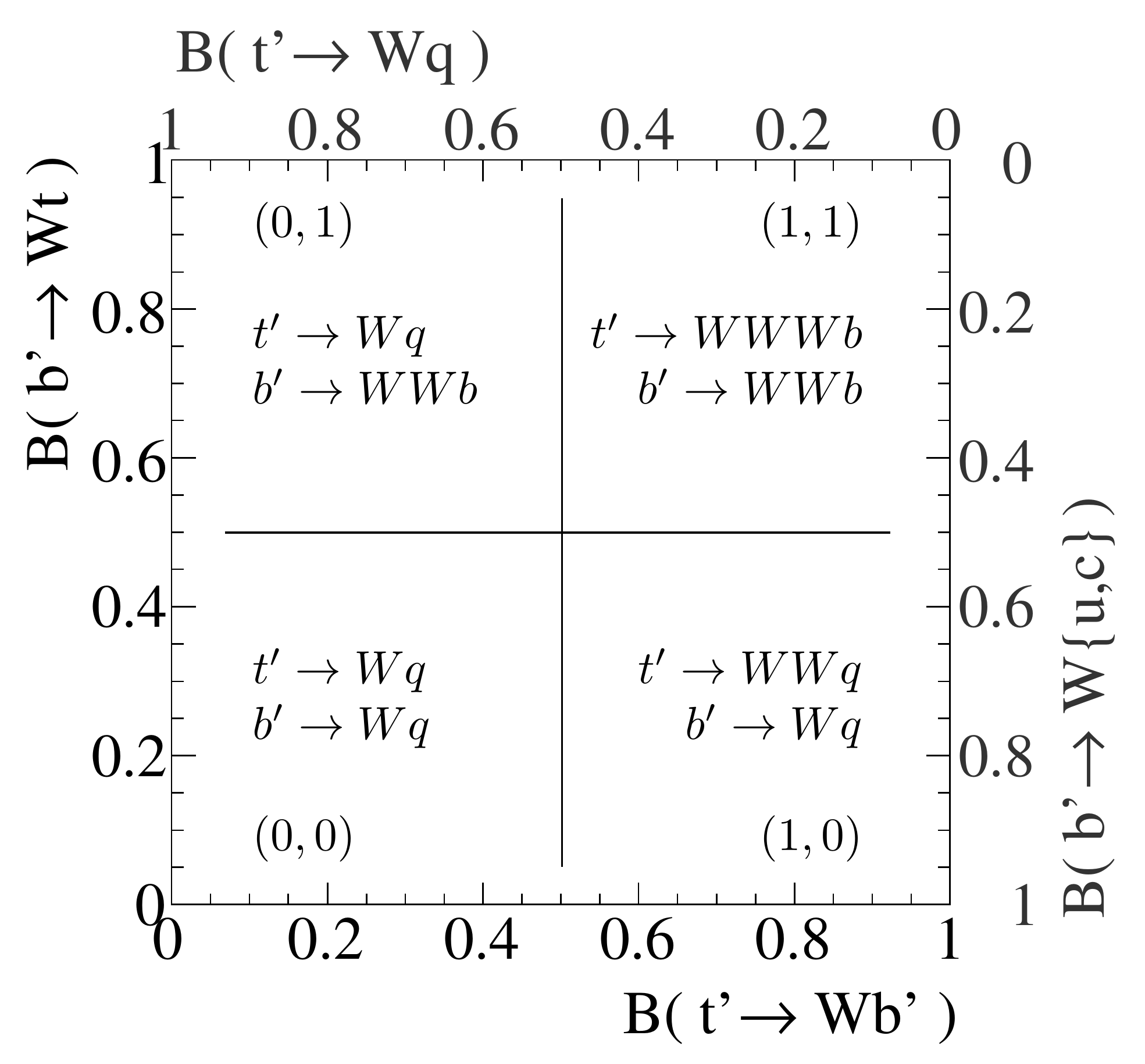}
\caption{ The flavor-mixing interval overlaid with a table of the processes contributing
to the corner points}
\label{fig:corners}
\end{figure}

The topologies of $b'$ and
$t'$ decays are determined by four branching fractions, two of which are independent:
\begin{eqnarray}
\mathcal{B}(t'\rightarrow Wb') & =  &1 - \mathcal{B}(t'\rightarrow W\{q=d,s,b\}) \nonumber \\
\mathcal{B}(b'\rightarrow Wt) & = & 1 - \mathcal{B}(b'\rightarrow W\{q=u,c\}) \nonumber
\end{eqnarray}
The dependence among these quantities, and the
processes they represent, is shown in Fig.~\ref{fig:corners}. In this representation,
the \lfj analysis, \lljbme and \lfivej analyses probe the corner $(0, 1)$.
Each used the assumption of an individual contribution from one flavor
of fourth-generation quark. 

We consider the implications of the CDF results in flavor-doublet scenarios,
characterized by different assumed mass splittings and a continuum flavor-mixing rates. 
To extend the interpretations of the published results, we use 
the basic relationship among event yield, cross section and acceptance to reinterpret the
observed yield limits under different assumptions. This requires careful estimation of the relative
acceptance between the original assumptions, and those of the new interpretation.

%We analyze individually the boundary cases, $[\mathcal{B}(t'\rightarrow Wb'),\mathcal{B}(b'\rightarrow Wt) ] =
%(0,0) (0,1) (1,0)$ and $(1,1)$ (illustrated in Fig.~\ref{fig:corners}), then allow the branching fractions to vary from  zero to one.

\section{Mass Limits}

\begin{figure}[ht]
\includegraphics[width=0.48\linewidth]{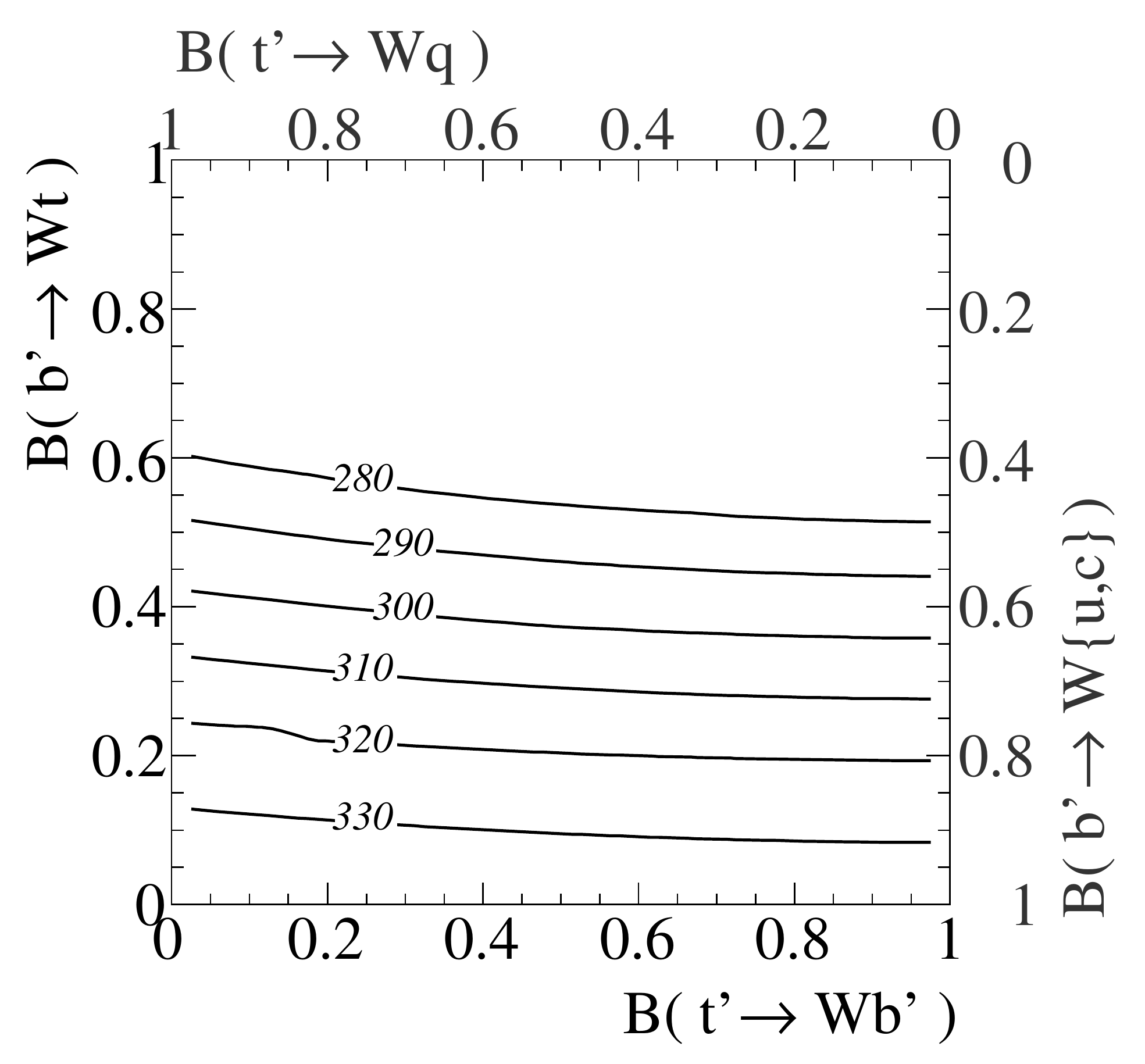}
\includegraphics[width=0.48\linewidth]{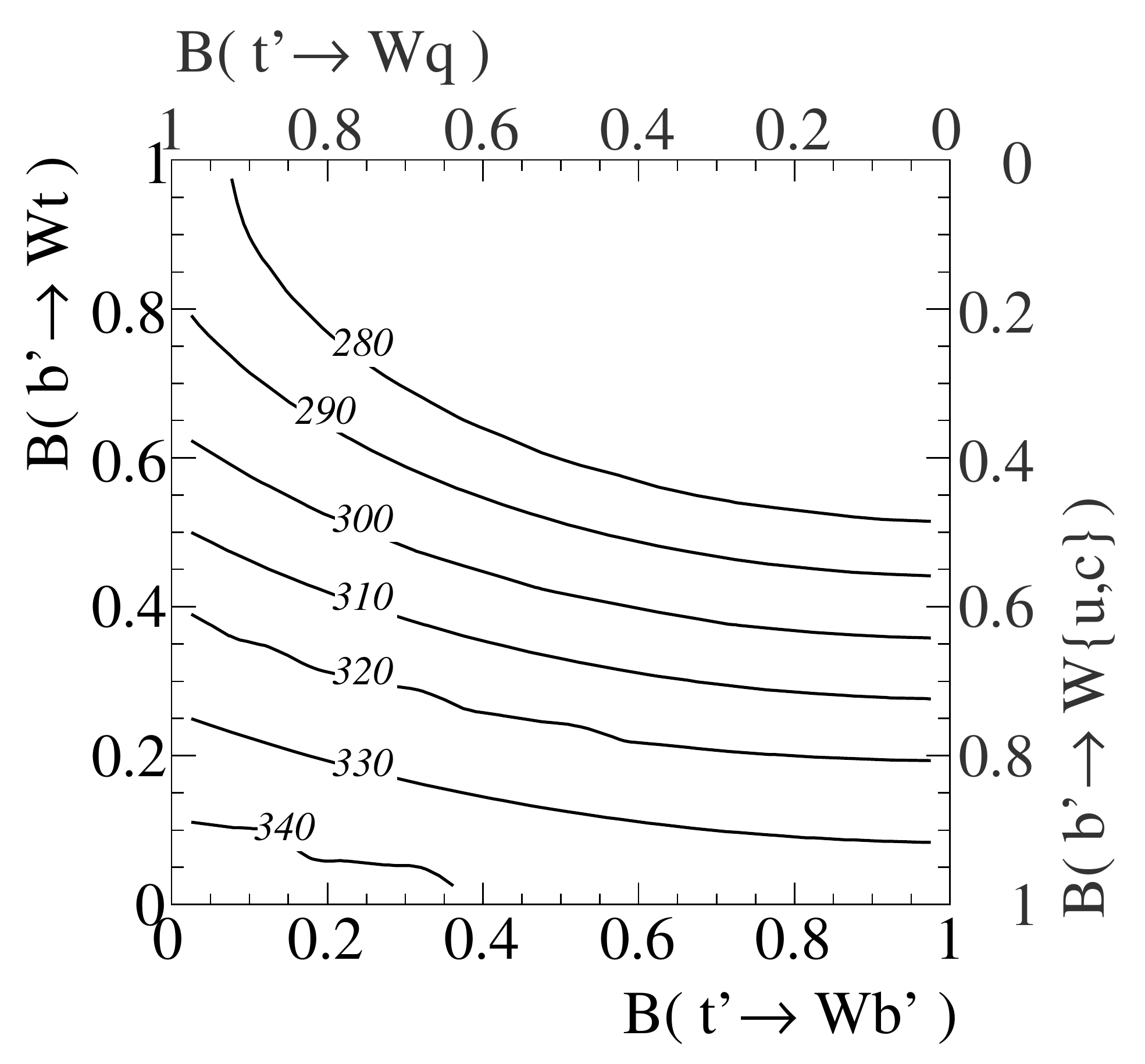}
\caption{ Limits on $b'$ mass from $\ell+4j$ data, as a function
  of branching fractions $\mathcal{B}(t'\rightarrow Wb') = 1 - \mathcal{B}(t'\rightarrow
  Wq)$ and  $\mathcal{B}(b'\rightarrow Wt') = 1 - \mathcal{B}(b'\rightarrow
  Wq)$ for mass structures $m_{t'} = m_{b'}+100\textrm{GeV}$ (left) and $m_{t'} =
  m_{b'}+50\textrm{GeV}$ (right).}
\label{fig:wq}
\end{figure}

\begin{figure}[ht]
\includegraphics[width=0.48\linewidth]{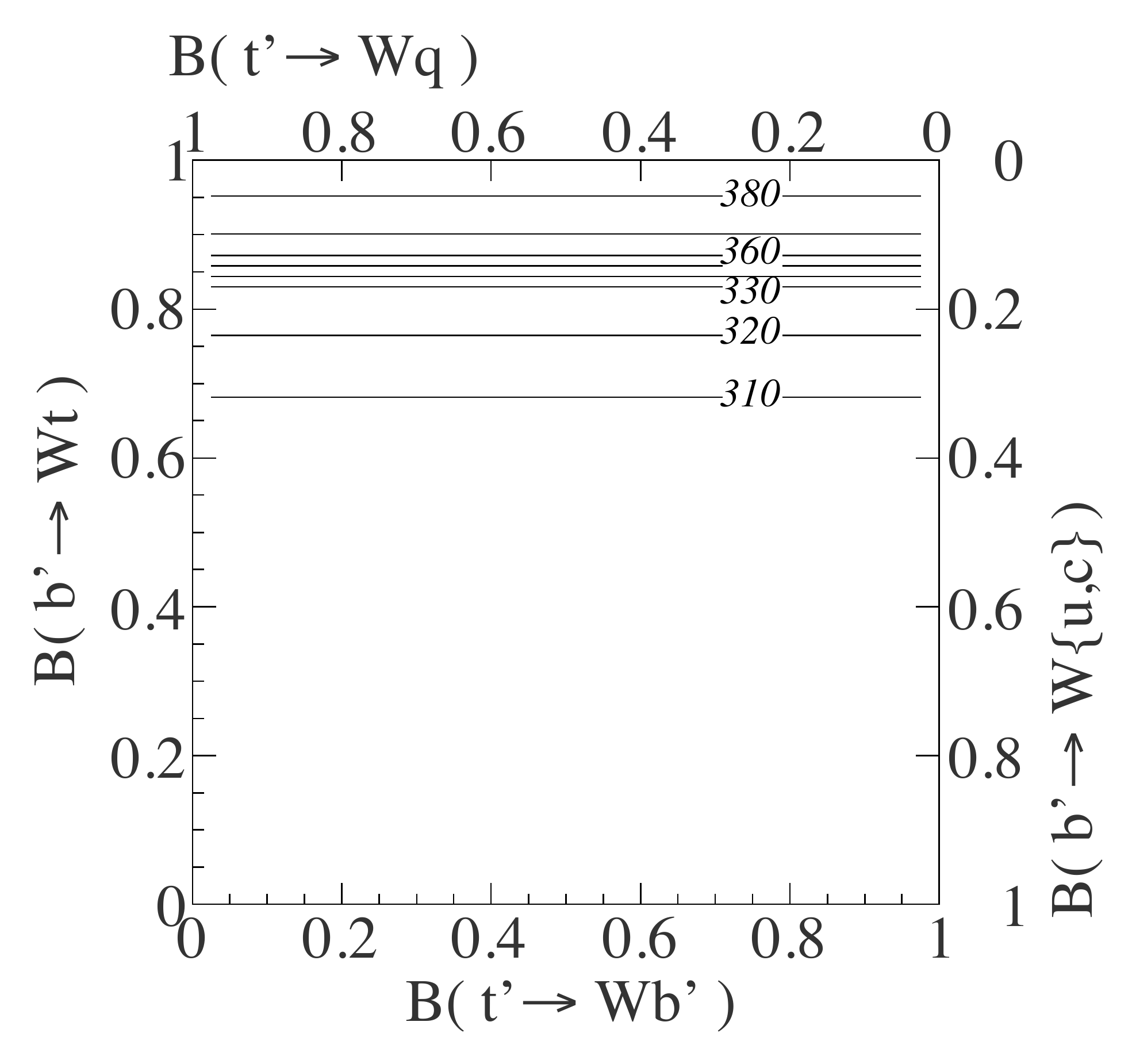}
\caption{ 
  Limits on $b'$ mass from \lfivej data, as a function
  of branching fractions $\mathcal{B}(t'\rightarrow Wb') = 1 - \mathcal{B}(t'\rightarrow
  Wq)$ and  $\mathcal{B}(b'\rightarrow Wt') = 1 - \mathcal{B}(b'\rightarrow
  Wq)$. The limits do not depend on $m_{t'} - m_{b'}$ as $t'$ gives no contribution.}
\label{fig:lfivej}
\end{figure}

\begin{figure}[ht]
\includegraphics[width=0.48\linewidth]{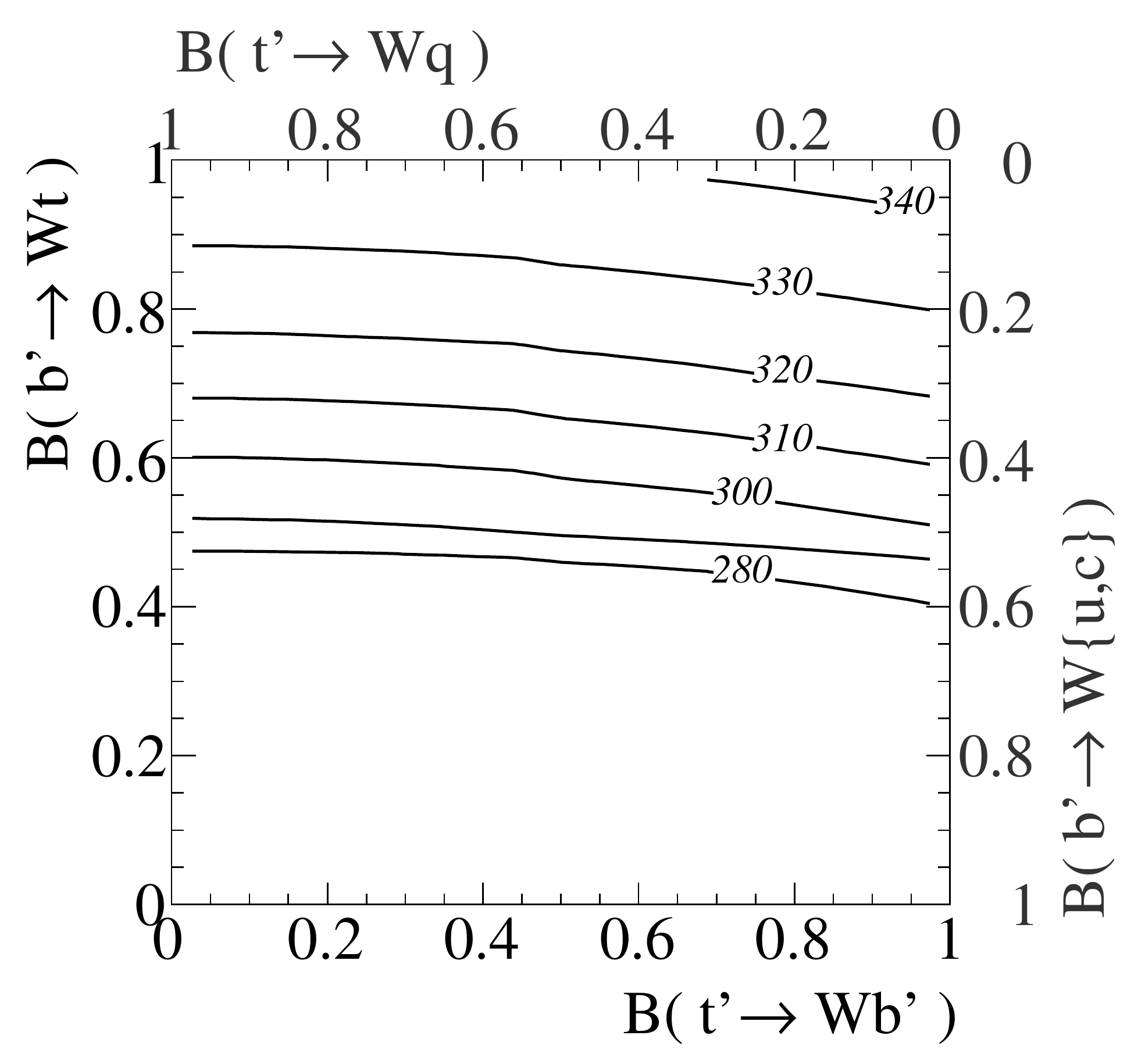}
\includegraphics[width=0.48\linewidth]{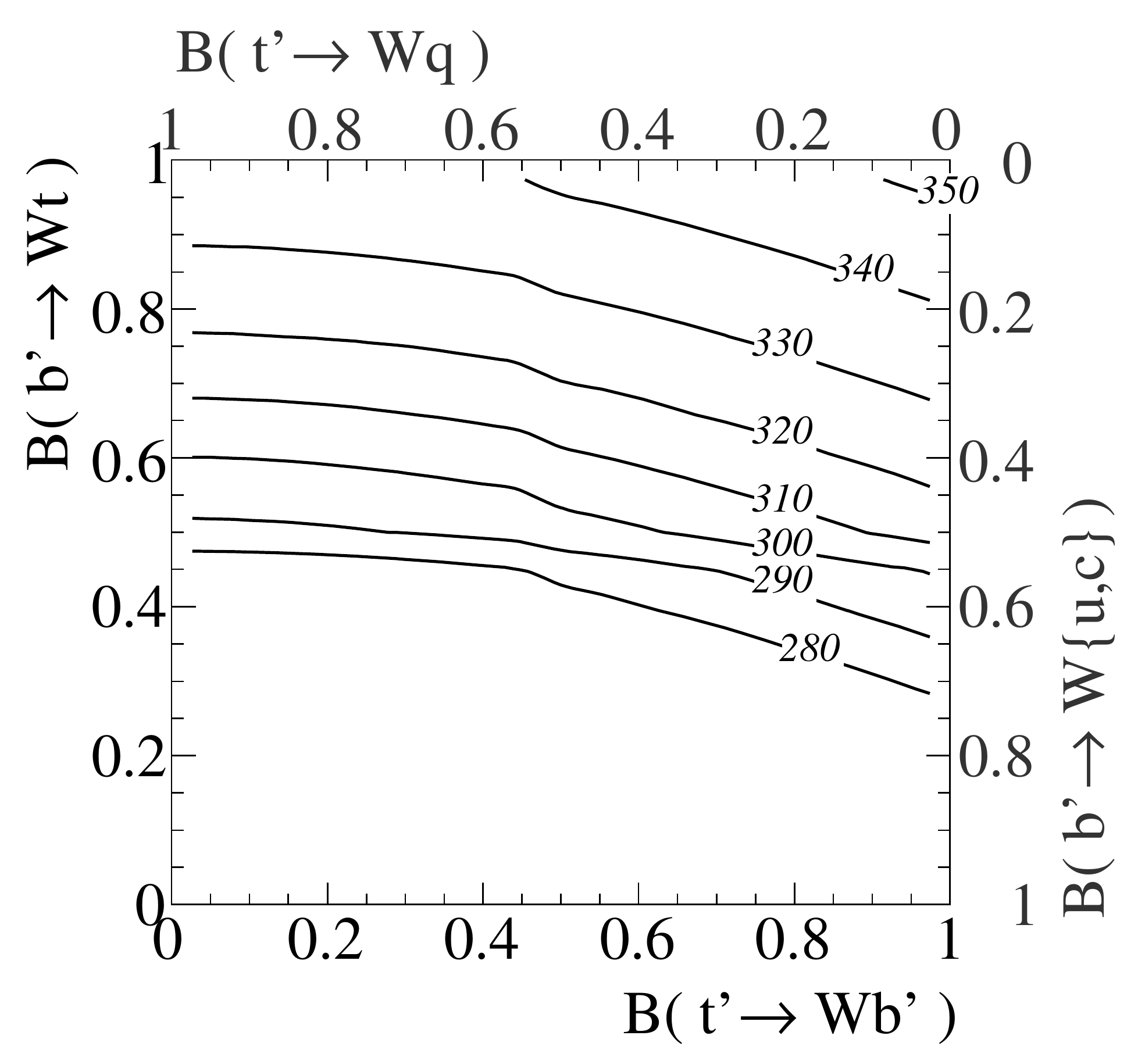}
\caption{ Limits on $b'$ mass from $\ell^{\pm}\ell^{\pm}jb\missET$ data, as a function
  of branching fractions $\mathcal{B}(t'\rightarrow Wb') = 1 - \mathcal{B}(t'\rightarrow
  Wq)$ and  $\mathcal{B}(b'\rightarrow Wt') = 1 - \mathcal{B}(b'\rightarrow
  Wq)$ for mass structures $m_{t'} = m_{b'}+100\textrm{GeV}$ (left) and $m_{t'} =
  m_{b'}+50\textrm{GeV}$ (right).}
\label{fig:wwb}
\end{figure}

\begin{figure}[ht]
\includegraphics[width=0.9\linewidth]{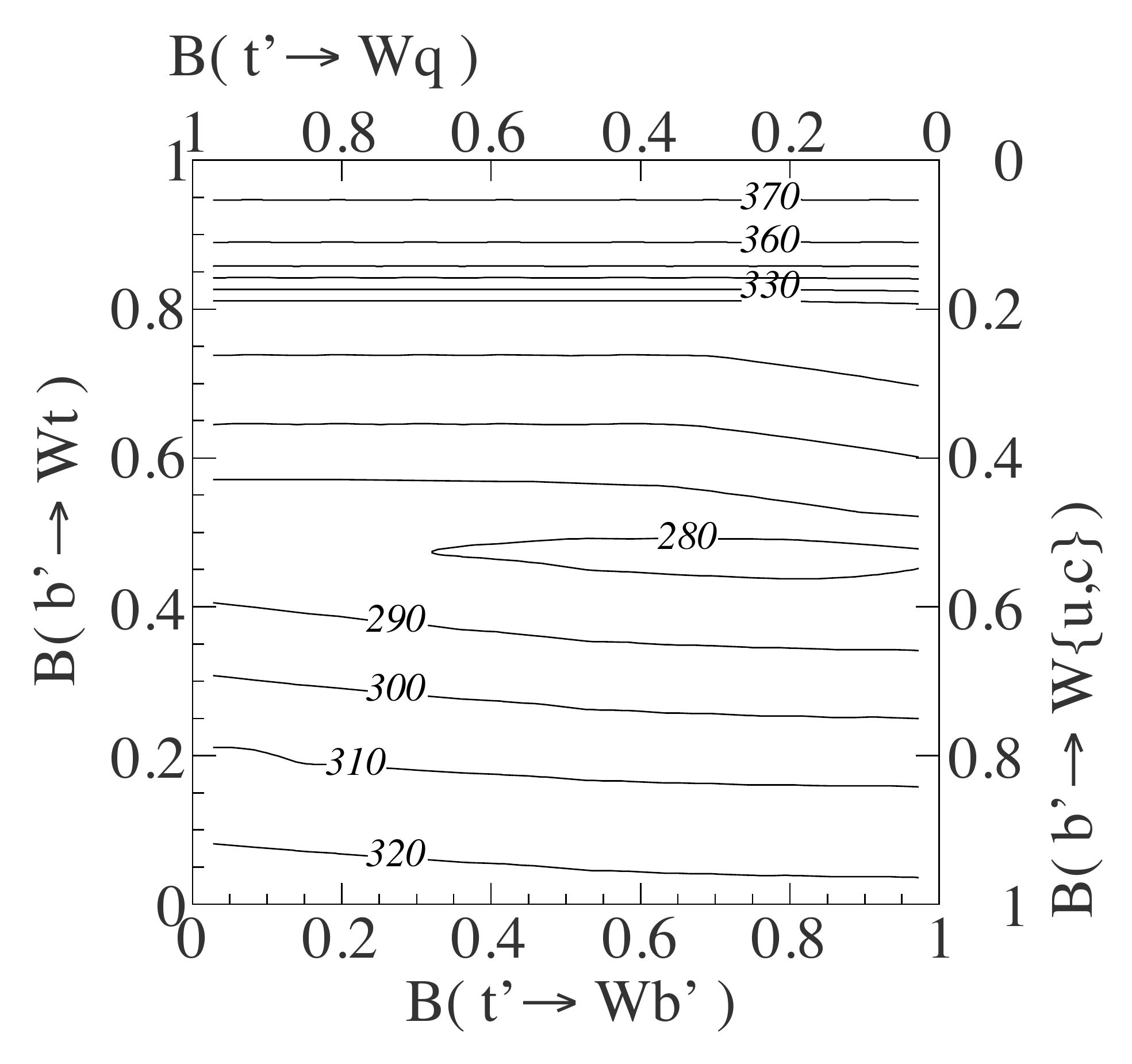}\\
\includegraphics[width=0.9\linewidth]{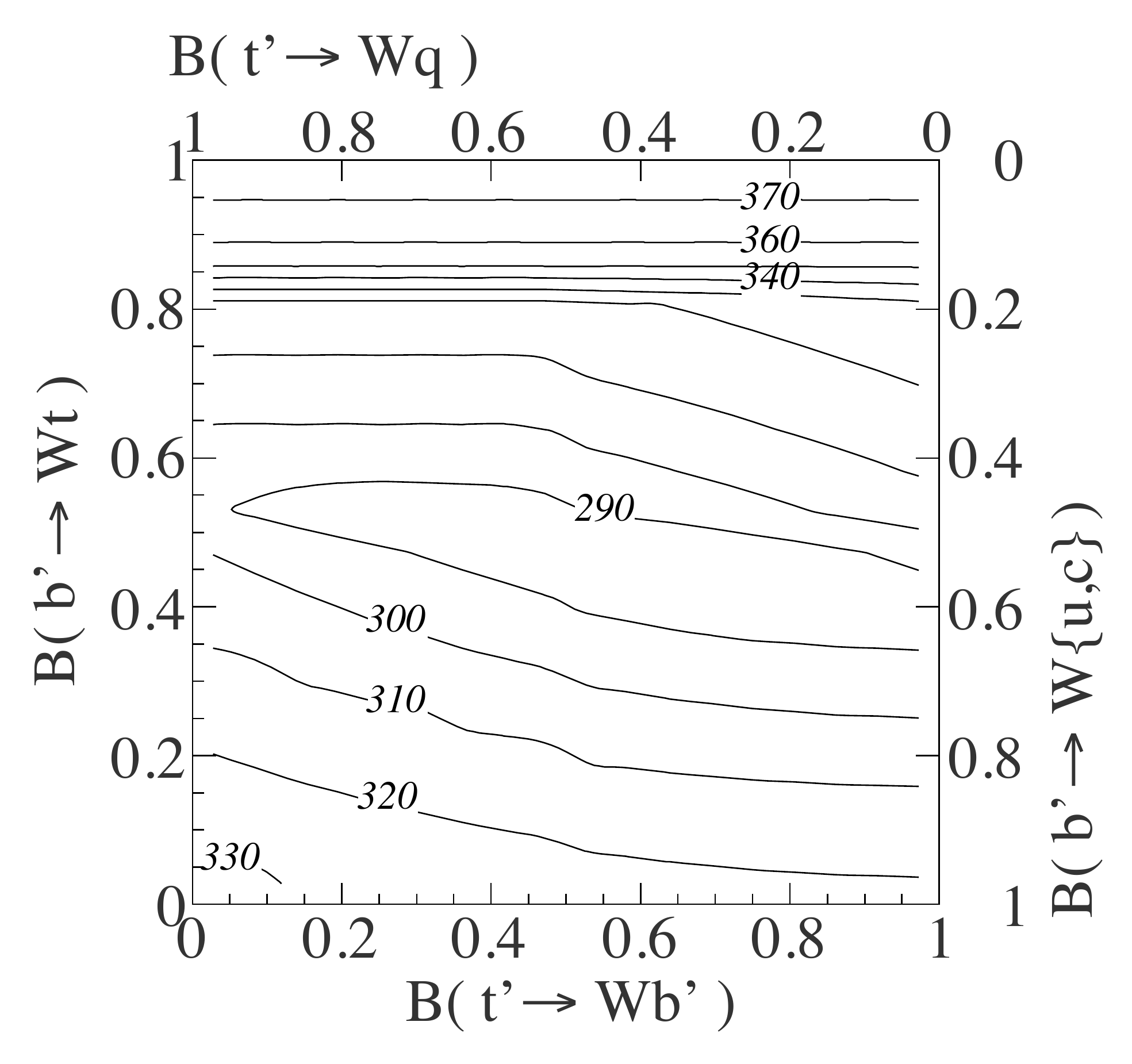}
\caption{ 
  Limits on $b'$ mass from $\ell^{\pm}\ell^{\pm}jb\missET$, \lfivej and
  $\ell+4j$  data, as a function
  of branching fractions $\mathcal{B}(t'\rightarrow Wb') = 1 - \mathcal{B}(t'\rightarrow
  Wq)$ and  $\mathcal{B}(b'\rightarrow Wt') = 1 - \mathcal{B}(b'\rightarrow
  Wq)$ for mass structures $m_{t'} = m_{b'}+100\textrm{GeV}$ (top) and $m_{t'} =
  m_{b'}+50\textrm{GeV}$ (bottom).}
\label{fig:wqwwb}
\end{figure}

%First, we consider the \lljbme sample, interpreted under the assumption of a \bprime and 
%\tprime with a classical mass splitting. In the original \lljbme interpretation, the event yield was 
%assumed to come from an individual \bprime . However, this yield could be interpreted under a model 
%with a \tprime contribution. We consider such a model, with decay modes of interest 
%\begin{eqnarray}
%b' & \rightarrow & Wt \rightarrow WWb \nonumber \\
%t' & \rightarrow & Wb'\rightarrow WWt \rightarrow WWWb \nonumber
%\end{eqnarray}
%The latter corresponds to the boundary case (1,1) in branching fraction space.
% The $t' \rightarrow WWWb$ mode has no current 
%direct limit even though 
%it would have a similar signature in the
%$\ell^{\pm}\ell^{\pm}jb\missET$ dataset, with larger acceptance due to the two additional
%$W$s in the intermediate decay chain. Indeed, if both fourth-generation 
%quarks exist, we would expect to select both
%modes in the $\ell^{\pm}\ell^{\pm}jb\missET$ sample.

In general, the event yield divided by the integrated luminosity, 
$N/L$, is equal to the cross section times the acceptance rate. For
a particular process, such as individual \bprime, this simply tells us that
the limit on a cross section, given an observed yield limit, is given by
\begin{equation}
\sigma_{\bprime} = \frac{N} {L \cdot \epsilon_{\bprime}}\textrm{,}
\end{equation}
where $\epsilon_{\bprime}$ is the acceptance rate for the observed process
within the experimental selection constraints. However, we can also consider
the case with two contributions if we know the relative acceptance rates between the 
processes, $\epsilon_{rel}$, and if the two cross sections are dependent:
\begin{equation}
\frac{N} { (L\cdot\epsilon_{\bprime})} =  \sigma_{\bprime} + \epsilon_{rel}\cdot\sigma_{\tprime}(\footnotesize\sigma_{\bprime}).
\end{equation}

Here, the dependence of the \tprime cross section on the \bprime cross section is given,
from the next-to-leading order cross-section calculations for massive quarks\cite{xsnlo} ,
by the mass splitting. \emph{Need to add comment on sigma rel!}
%The published theoretical cross-section was fit to an exponential 
%model in the local region under consideration. The decay constant $M$ was found to be
%$40$GeV.
%\begin{equation}
%\sigma_{\tprime} = \sigma_{\bprime}e^{-\frac{m_{\tprime}-m_{\bprime}}{M}},\ \sigma_{rel} = e^{-\frac{m_{\tprime}-m_{\bprime}}{M}}
%\end{equation}
%We use this relationship
%to produce cross-section limits, and by extension mass limits, 
%on \bprime for a spectrum of assumed \tprime 
%masses. The relative efficiency  $\epsilon_{rel}$ between the original model
%and the model explicitly including the \tprime, was 
%estimated using simulated data. 
%The model for $\epsilon_{rel}$ accounts for increased acceptance of \tprime 
%as the fourth-generation mass difference
%increases, with a plateau beyond $m_W$. This method was validated by 
%varying the model for $\epsilon_{rel}$, and the resulting mass limits were found to be stable.

%In the same way, we calculate limits from the \lljbme sample at the corner $(0, 0)$
%in the two-flavor case. We choose mass splititngs of 50 and 100GeV and find limits
%of $m_{\bprime}>350$ and XXX GeV respectively.

To probe the full two-dimensional branching fraction interval, 
we must calculate the dependence of the event yield on the branching fractions
explicitly. As the branching fractions to the reconstructed states vary, the acceptances of the 
processes of interest vary accordingly. Reconstruction 
efficiencies can differ as well. Considering these effects, we calculate the acceptances of \bprime and \tprime
in the \lfj, \lfivej and \lljbme samples.

%We present results for two classical mass splittings: $m_{\tprime}-m_{\bprime}$ of 50GeV and 100GeV.

The expected signal yield for one process relative to another is proportional to relative production rates and final-state reconstruction efficiencies. The relative signal production rates have two  factors:
the relative 
cross section for initial state production and the branching ratios of the involved processes (second-order when both sides of the event are considered, as here). When several processes contribute to a signal there are multiple terms of this form. Fixing the event
yield at its observed value, we may isolate the cross section we wish to limit, and express it in terms of the previously
measured limit and an effective relative acceptance. The effective relative acceptance $A$ includes acceptance terms for all processes
considered, each scaled by relative cross-section.

In the \lfj case, there are no relative reconstruction efficiencies to consider, and we have a simple
expression for the relative expected yields as a function of $\beta_{b'} = \mathcal{B}(b'\rightarrow Wt)$ and
$\beta_{t'}=\mathcal{B}(t'\rightarrow Wb')$:
\begin{equation}
 A(\beta_{b'},\beta_{t'}) = (1-\beta_{b'})^2 + \sigma_{rel}(1-\beta_{t'})^2  
\end{equation}
We use this expression to produce limits on the mass of the \bprime as a function of fourth-generation
branching fractions. The results, shown in the interval introduced before, are presented in Fig.~\ref{fig:wq}.

The \lfivej case is similar, as we are limited in our reinterpretation
of the published results, and can only write

\begin{equation}
 A(\beta_{b'},\beta_{t'}) = \beta_{b'}^2 
\end{equation}

with the limits shown in Fig.~\ref{fig:lfivej}.

The \lljbme case is somewhat more complicated. In addition to \bprime, there are two significant \tprime
 processes that produce the signal selected for the sample analyzed:
 \begin{eqnarray}
  t'\rightarrow Wb' & \rightarrow & WWt \rightarrow WWWb \nonumber \\
  t'\rightarrow Wb' & \rightarrow & WWc \nonumber
 \end{eqnarray}
 Moreover, because of jet multiplicity and the flavor-tag requirement, it is necessary to consider relative reconstruction efficiencies for each contribution. These factors are denoted
 $\epsilon_{NW}$, where $N$ is the number of intermediate $W$-bosons in the process described. Presented in Table \ref{effs}, they are estimated from
 simulated data and are calculated relative to the four-$W$ case considered in the original analysis.
\begin{eqnarray}
\lefteqn{A(\beta_{b'},\beta_{t'}) = \beta_{\bprime}^{2} + \beta_{\tprime}^{2}\frac{ \sigma_{rel}}{\epsilon_{bb}} [  (1-\beta_{b'})^2\epsilon_{4W} \epsilon_{cc}} \nonumber \\
 & + & 2\beta_{b'}(1-\beta_{b'})\epsilon_{5W}\epsilon_{cb} + \beta_{b'}^2\epsilon_{6W}\epsilon_{bb}]
\end{eqnarray}
The different terms in the \tprime contribution have different reconstruction efficiencies due to jet-flavor tagging, expressed by
$\epsilon_{f_1f_2}$ where $f_1f_2$ is the flavor combination of the quarks in the final state of the quark-level \tprime decay.
These efficiencies are given by statistics from the raw efficiencies of the jet-flavor tag to select the beauty (60\%)
or charm (15\%) flavor in a jet:
\begin{eqnarray}
\epsilon_{bb} & = & 1 - (1-\epsilon_b)^2 \nonumber \\
\epsilon_{cb} & = & \epsilon_c(1-\epsilon_b) +\epsilon_b(1-\epsilon_c) + \epsilon_b\epsilon_c \nonumber \\
\epsilon_{cc} & = & 1 - (1-\epsilon_c)^2 \nonumber
\end{eqnarray}
Again, we produce limits on the mass of the \bprime as a function of fourth-generation
branching fractions. The results are presented in Fig.~\ref{fig:wwb}.

As expected, inspection of these results reveals the complementary sensitivities of the CDF analyses.
However, they are not orthogonal and thus cannot be statistically combined. In view of this, we 
produce the best limits over the branching fraction interval by choosing the stronger of two limits
at each point. The combined results are presented in Fig.~\ref{fig:wqwwb}.

\begin{table}
\caption{Reconstruction efficiencies for $t'$ decays mediated by W bosons, relative to the $b'$
reconstruction efficiency of the original CDF analysis.}

\begin{tabular}{ c | c | c }

$m_{t'} - m_{b'}$ & 50GeV & 100GeV \\
\hline \hline
$\epsilon_{4W}$ & 0.63 & 0.86 \\
\hline
$\epsilon_{5W}$ & 1.07 & 1.51 \\
\hline
$\epsilon_{6W}$ & 1.60 & 2.16 \\

\end{tabular}

\label{effs}
\end{table}

\section{CKM4 Mapping}

To compare results of many different experiments, it is useful to
decouple the various physical parameters wherever possible and express the results in terms of the underlying
physical quantities at the theoretical level. In this case, these are the elements of the
flavor mixing matrix, on which there are important experimental constraints.

The description of SM quark flavor mixing falls to the CKM matrix, which, in 
the presence of a fourth generation, is extended to CKM4. It has 16 elements parametrized by 4 mixing angles and three irreducible 
CP-violating phases\cite{ckm4}.  Present experimental data has been brought to
bear on the extent to which this could be consistent with observation (cf. \cite{ckm4}). 

The ideal case would be to form a clear basis of comparison between theory and experiment, and some of the challenges of
doing that are topics of discussion here. The primary difficulty is in decoupling the many
dependent parameters involved, so one aims to formulate analysis to
this end.  Here, we eliminate many degrees of freedom by treating light quarks as indistinguishable.
The scale of light quark mass is relative to the mass difference with $W$, as required for weak decay.
This treatment is also natural from the perspective of experimental
design.    

In the limit of massless particles, the branching fractions of the previous sections translate directly to 
the squares of the magnitudes of the corresponding CKM4 elements:

\begin{eqnarray}
\mathcal{B}(t'\rightarrow Wb') &=& |V_{t'b'}|^2 \nonumber \\
\mathcal{B}(b'\rightarrow Wt) &=& |V_{tb'}|^2 \nonumber
\end{eqnarray}

In the case of massive particles, the partial widths carry on this proportionality,
but there are also phase-space constraints to consider. The branching fraction of the 
massive particle is of course constructed from the partial widths of the various possible
modes. Thus the branching fraction has dependence on the masses of the parent
and daughter particles.

This presents a challenge: can we discuss fourth generation mass limits
in the context of a CKM parametrization? Ignoring for the moment 
the issue of mass dependance, we can construct
a transformation between the branching fractions and the CKM
elements.

The plane we have chosen, with axes $\mathcal{B}(t'\rightarrow Wb') = 1 - \mathcal{B}(t'\rightarrow
Wq)$ [$q=d,s,b$] and  $\mathcal{B}(b'\rightarrow Wt) = 1 - \mathcal{B}(b'\rightarrow
Wq)$ [$q=u,c$] includes processes with tree-level diagrams having vertex factors of $V_{44}$ and $V_{34}$
respectively. Of course, a branching fraction is the quotient ${\Gamma_{\textrm{partial}}} / {\Gamma_{\textrm{total}}}$.
  
At tree-level, the partial widths of the final states are the product of phase-space and weak-coupling 
factors, and the absolute square of the tree-level CKM vertex factor.

For the total widths of the denominators, we would like to eliminate dependence on the mixing
angles between the fourth generation and the lighter quarks to minimize the number of variables 
involved in the transformation. At tree-level, this seems again possible:

  \begin{enumerate}
  
  \item The phase-space factors of matrix elements to the various flavors are nearly identical for
  all light quarks, including $b$ when the mother quark is a $t'$.
    
  \item By weak universality we can eliminate the CKM parameters that describe the mixing with lighter quarks.
     
  \end{enumerate}

\subsection{$\mathcal{B}(t'\rightarrow Wb')$ as a function of $V_{44}$}

For the numerator:
\[ \|V_{44}\|^2\cdot\|\int^{(t',b')} dp^4(...)\|^2 \]
The integral over four-space is meant to signify the full matrix element for the process
indicated by superscript, excluding only the CKM factor. The matrix element is calculated using the 
software program BRI \cite{BRI}, in a configuration
that models no quark mixing (to decouple the CKM suppression).

For the denominator, we have such a term for each generation:

\begin{eqnarray}
%\[ 
\|V_{41}\|^2\cdot\|\int^{(t',d)} dp^4(...)\|^2 +
\|V_{42}\|^2\cdot\|\int ^{(t',s)}dp^4(...)\|^2 \nonumber \\ +  \|V_{43}\|^2\cdot\|\int^{(t',b)} dp^4(...)\|^2 +  \|V_{44}\|^2\cdot\|\int ^{(t',b')}dp^4(...)\|^2 \nonumber \\
 =\{ \|V_{41}\|^2 +  \|V_{42}\|^2  +  \|V_{43}\|^2 \}\cdot\|\int
 ^{(t',q)}dp^4(...)\|^2\nonumber \\ +  \|V_{44}\|^2\cdot\|\int^{(t',b')} dp^4(...)\|^2 \nonumber\\
 = \{ 1 - \|V_{44}\|^2 \}\cdot\|\int ^{(t',q)}dp^4(...)\|^2 
 \nonumber \\
+ \|V_{44}\|^2\cdot\|\int^{(t',b')} dp^4(...)\|^2  \nonumber
%\]
\end{eqnarray}

In the case of the branching fraction for unmixed weak fourth-generation decay, we obtain a simple transformation between the branching fraction and the 
corresponding CKM vertex factor.

\subsection{$\mathcal{B}(b'\rightarrow Wt)$ as a function of $V_{43}$ and $V_{44}$}

\begin{figure}
\includegraphics[width=0.48\linewidth]{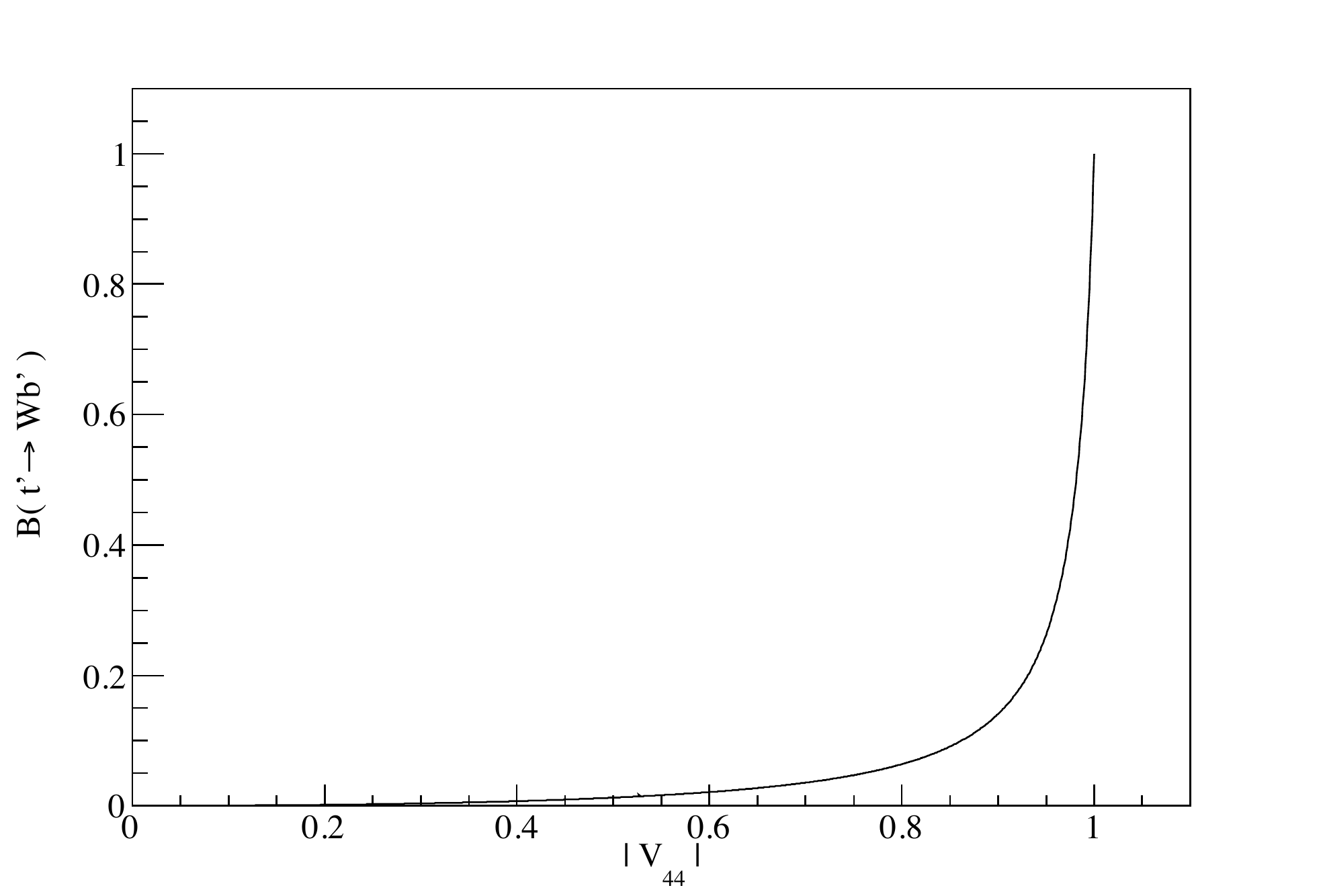}
\includegraphics[width=0.48\linewidth]{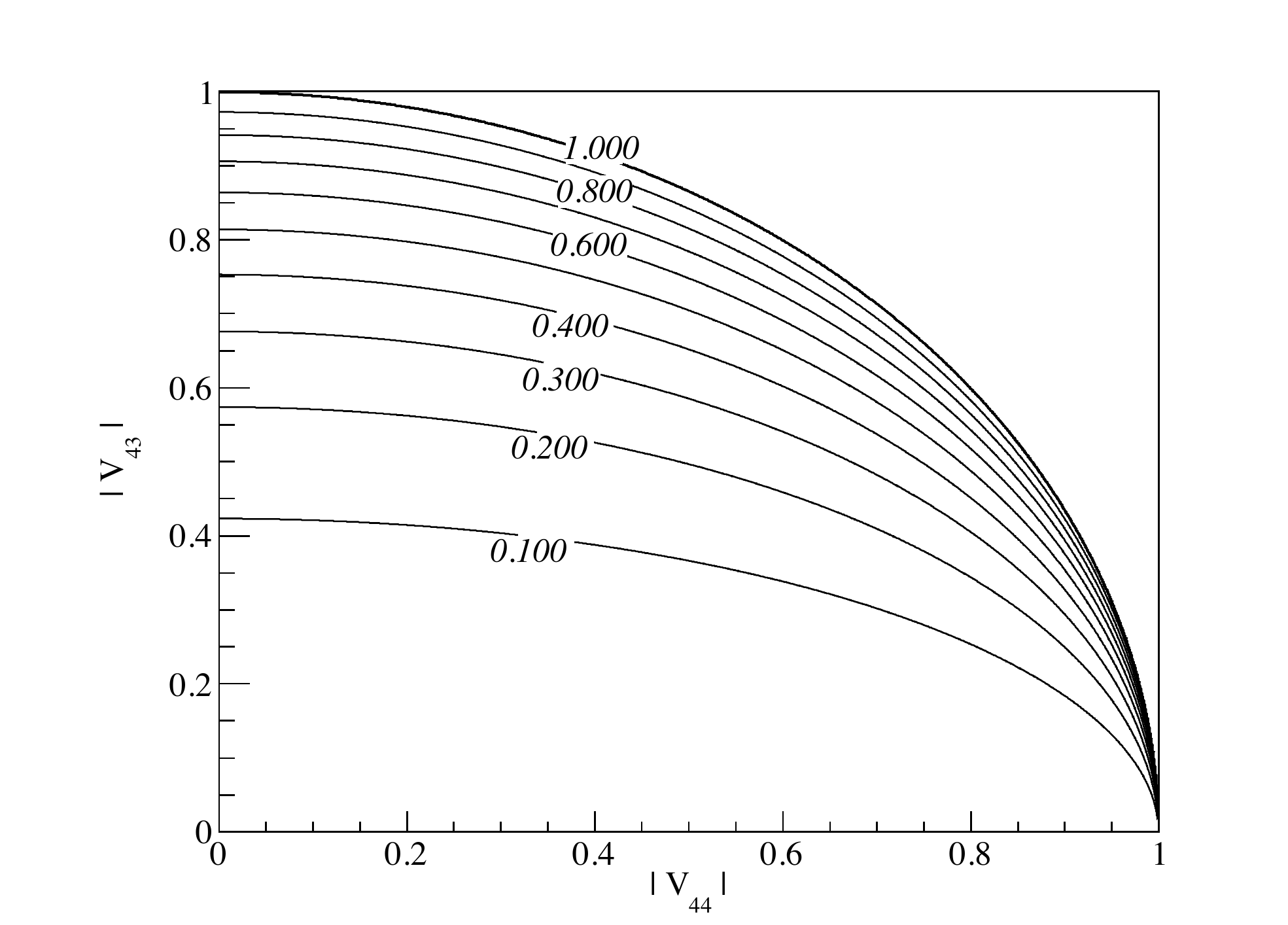}
\caption{ Mapping from branching fraction to
  CKM4 space for mass structure $m_{t'} = m_{b'}+100\ \textrm{GeV}$. Left, $\mathcal{B}(t'\rightarrow Wb')$ as a function of
  CKM4 parameters  $V_{44}$. Right,
  $\mathcal{B}(b'\rightarrow Wt)$ as a function of $V_{43}$ and $V_{44}$. }
\label{fig:CKMConst1}
\end{figure}

\begin{figure}
\includegraphics[width=0.48\linewidth]{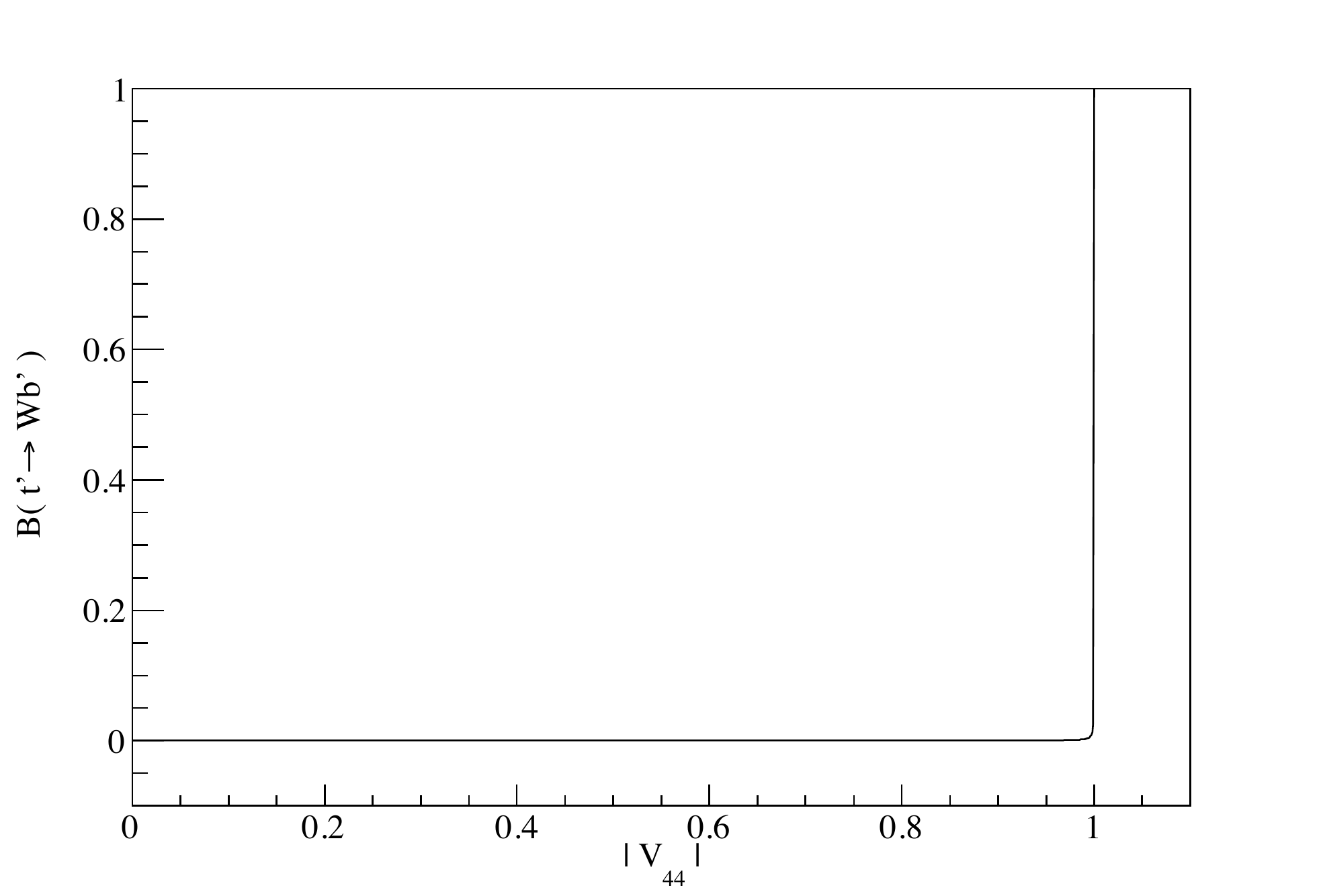}
\includegraphics[width=0.48\linewidth]{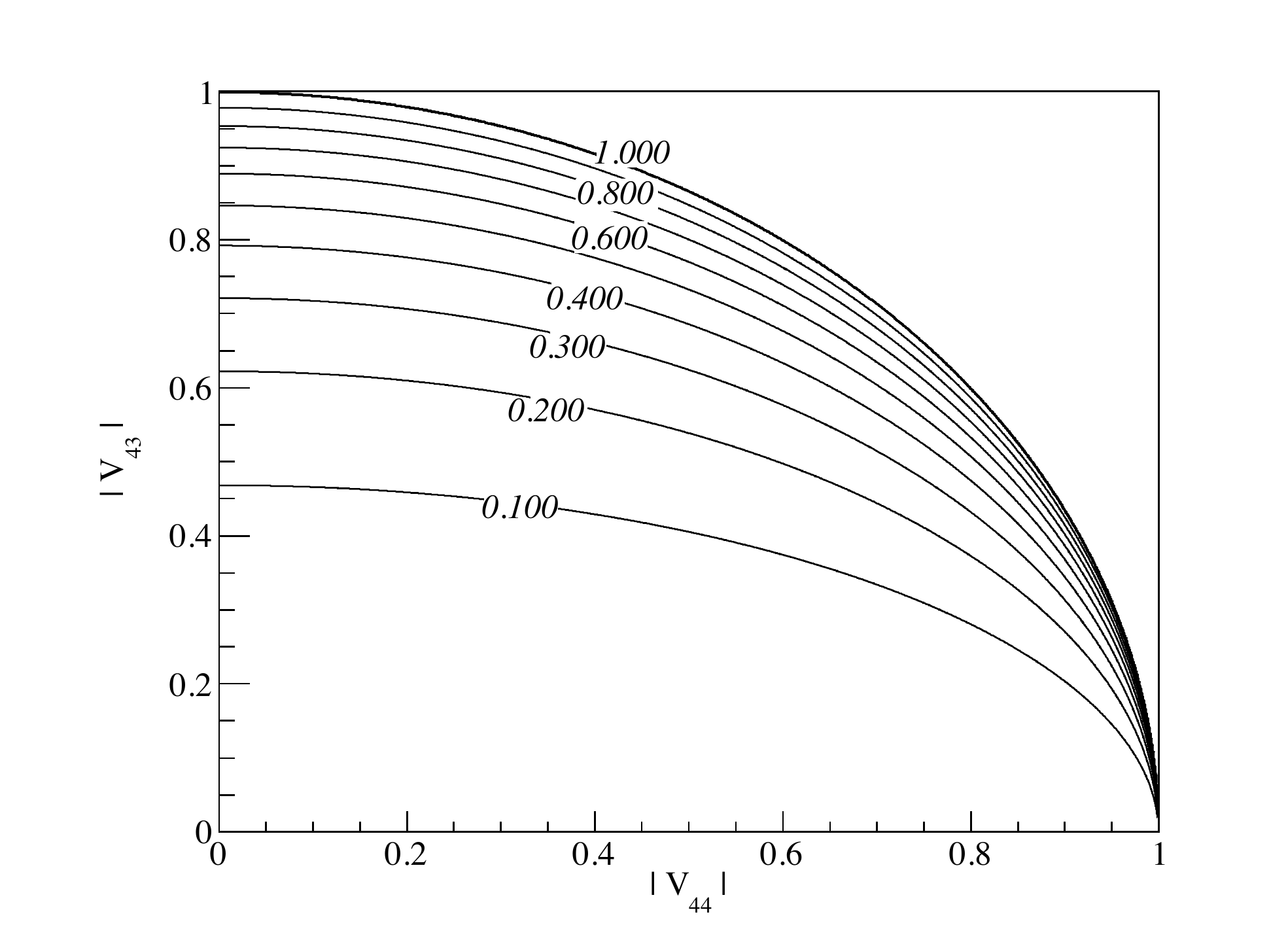}
\caption{  Mapping from branching fraction to
  CKM4 space for mass structure $m_{t'} = m_{b'}+50\ \textrm{GeV}$. Left, $\mathcal{B}(t'\rightarrow Wb')$ as a function of
  CKM4 parameters $V_{43}$ and $V_{44}$. Right,
  $\mathcal{B}(b'\rightarrow Wt)$. }
\label{fig:CKMConst1B}
\end{figure}

\begin{figure}
\includegraphics[width=0.9\linewidth]{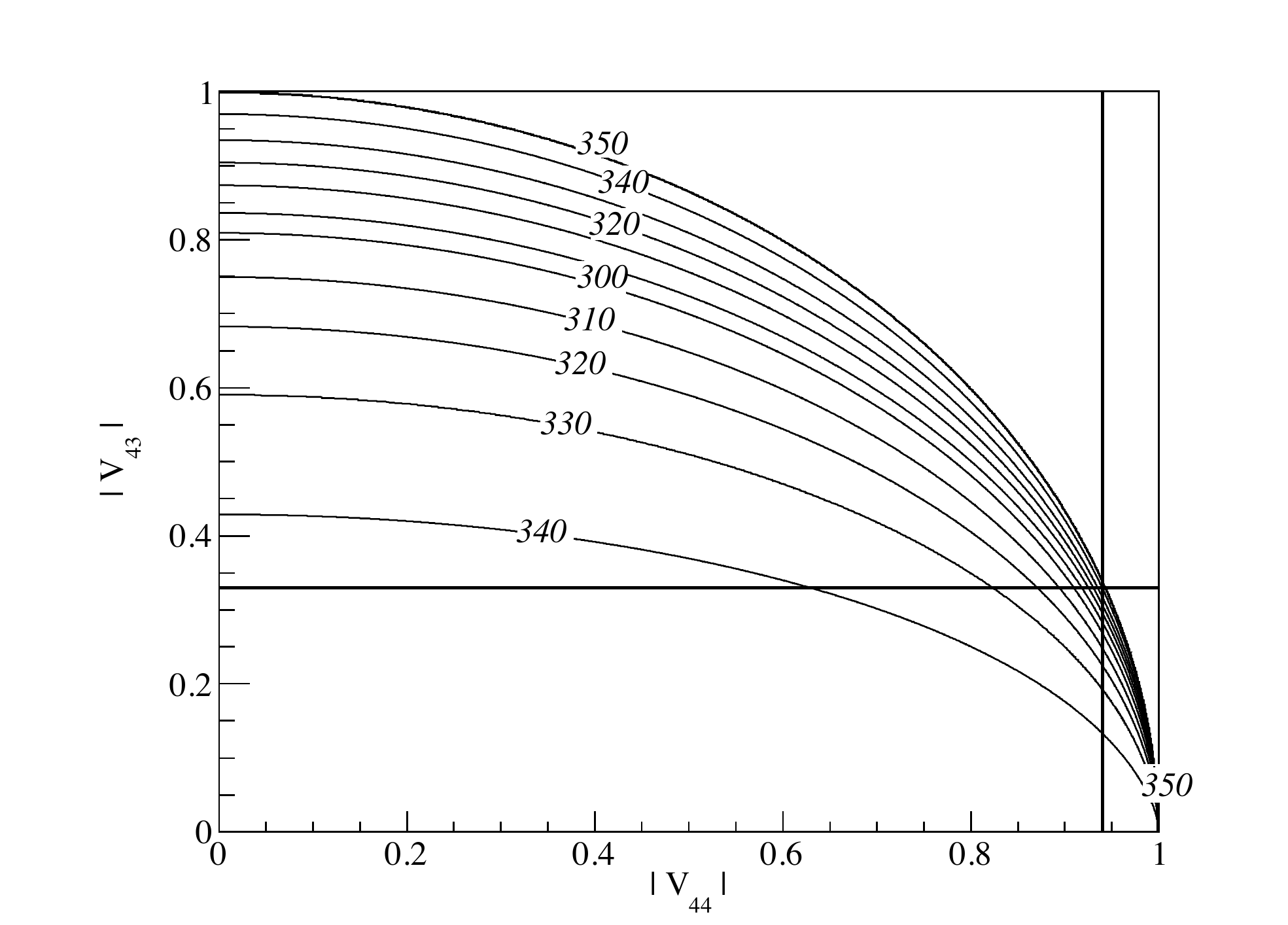}
\includegraphics[width=0.9\linewidth]{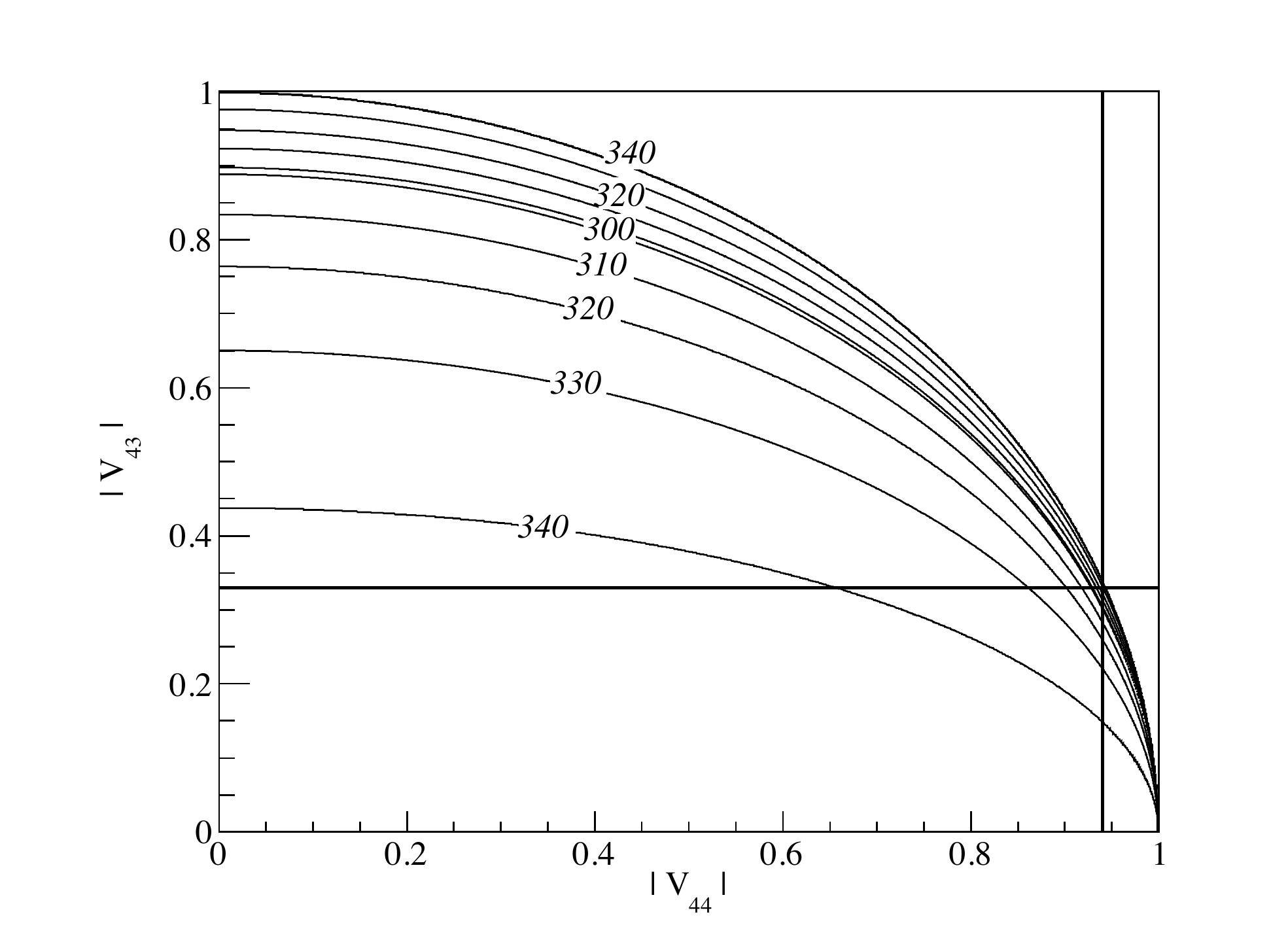}
\caption{ 
  Limits on $b'$ mass from $\ell^{\pm}\ell^{\pm}jb\missET$, \lfivej and
  $\ell+4j$  data, as a function of CKM4 parameters $V_{43}$ and $V_{44}$
  for mass structures $m_{t'} = m_{b'}+100\ \textrm{GeV}$ (top) and $m_{t'} =
  m_{b'}+50\ \textrm{GeV}$ (bottom).
 }
\label{fig:CKMConts2}
\end{figure}

\begin{figure}
\includegraphics[width=0.9\linewidth]{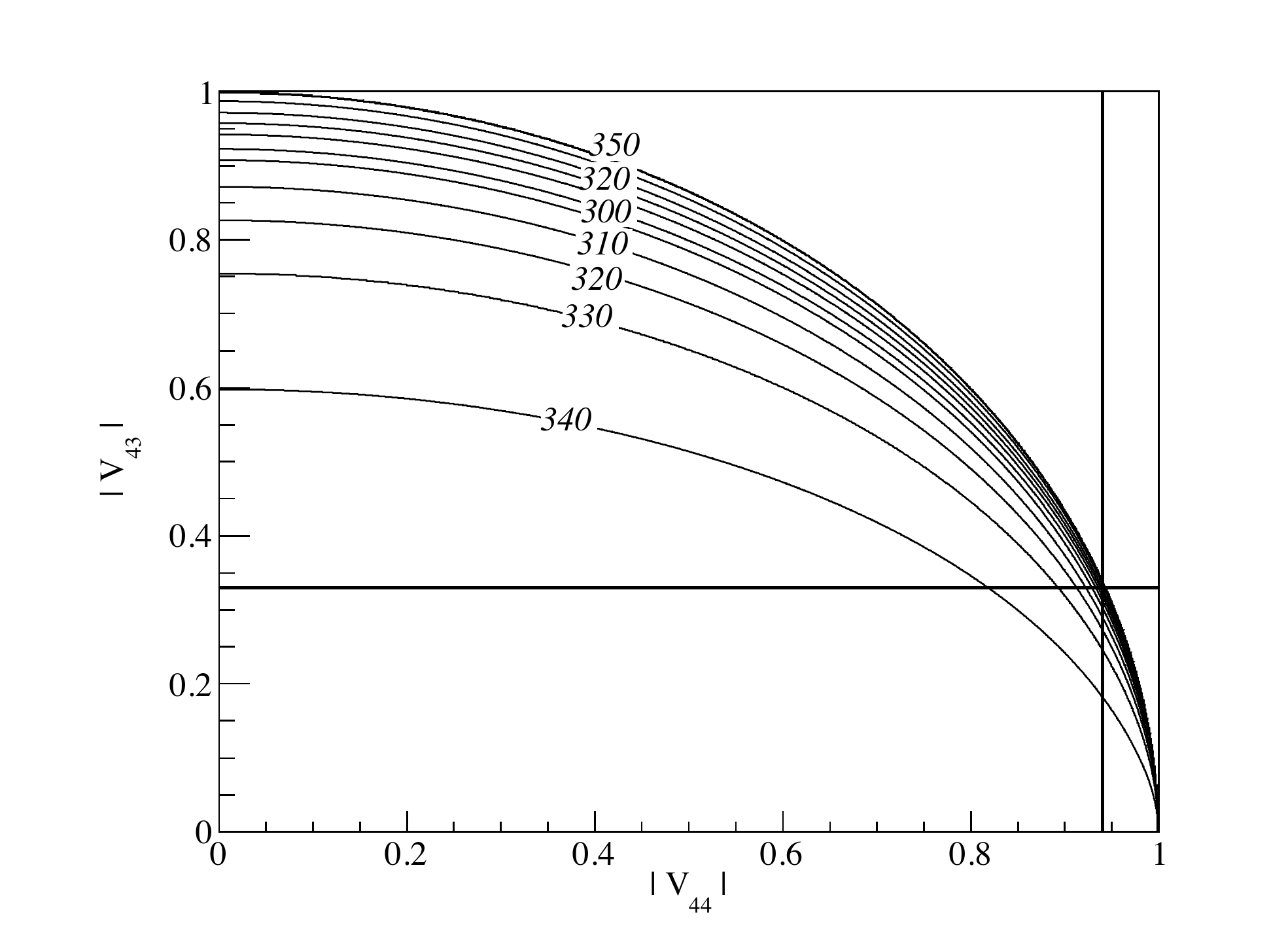}
\includegraphics[width=0.9\linewidth]{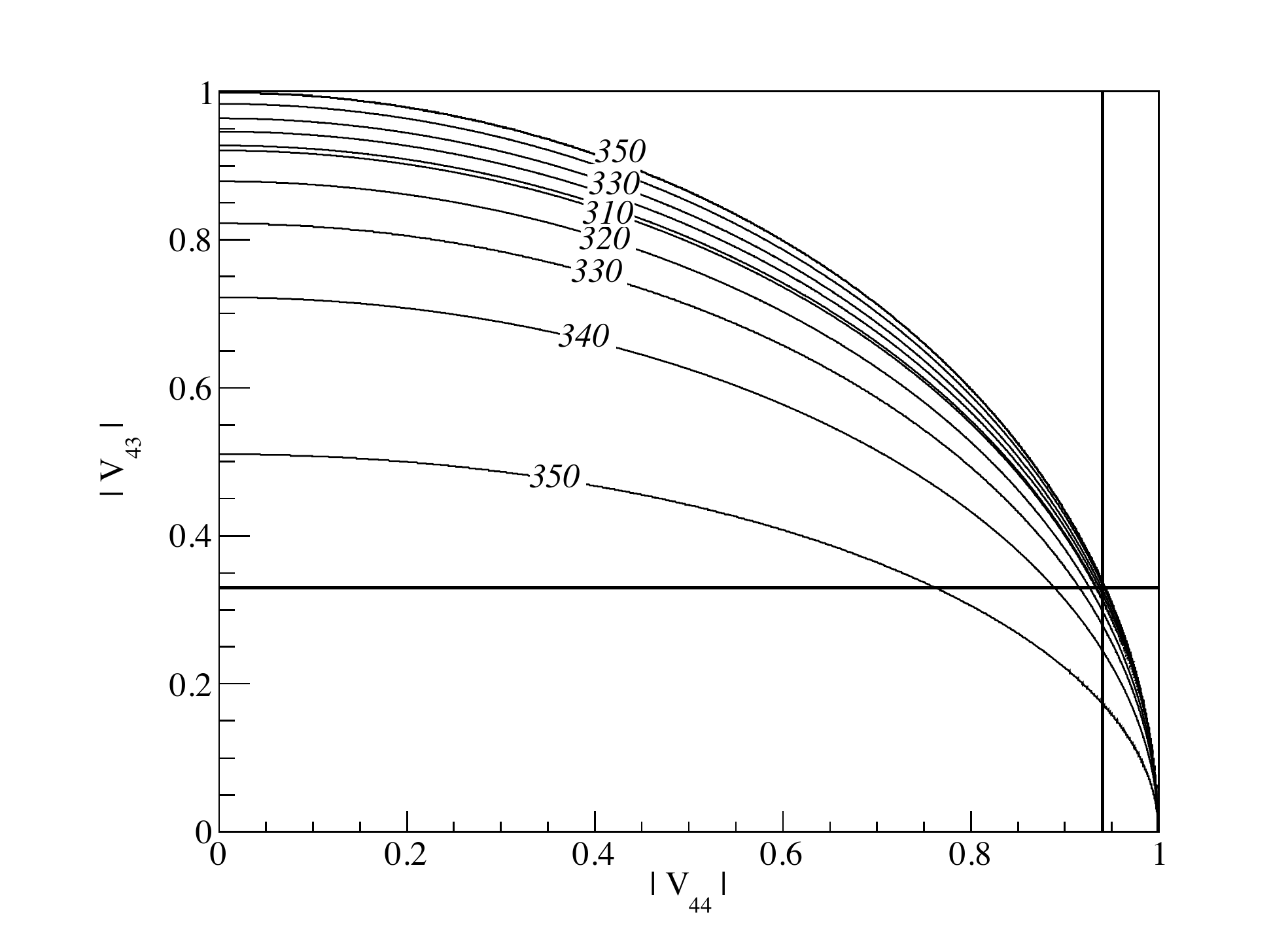}
\caption{ 
  Limits on $b'$ mass from $\ell^{\pm}\ell^{\pm}jb\missET$, \lfivej and
  $\ell+4j$  data, as a function of CKM4 parameters $V_{43}$ and $V_{44}$
  for mass structures $m_{t'} = m_{b'}+100\ \textrm{GeV}$ (top) 
  and  $m_{t'} = m_{b'}+50\ \textrm{GeV}$ (bottom) using
  alternate mappings from branching ratio space with varied
  phase-space assumptions. Compare to Fig.~\ref{fig:CKMConts2}.
 }
\label{fig:CKMConts2_alt1}
\end{figure}

%\begin{figure}[ht]
%\includegraphics[width=0.9\linewidth]{CKM_contours_m50_alt2}
%\caption{ 
 % Limits on $b'$ mass from $\ell^{\pm}\ell^{\pm}jb\missET$, \lfivej and
 % $\ell+4j$  data, as a function of CKM4 parameters $V_{43}$ and $V_{44}$
 % for mass structures $m_{t'} = m_{b'}+50\ \textrm{GeV}$ using an
 % alternate mapping from branching ratio space with varied
 % phase-space assumptions. Compare to Fig.~\ref{fig:CKMConts2} (bottom).
% }
%\label{fig:CKMConts2_alt2}
%\end{figure}

\begin{figure}
\includegraphics[width=0.9\linewidth]{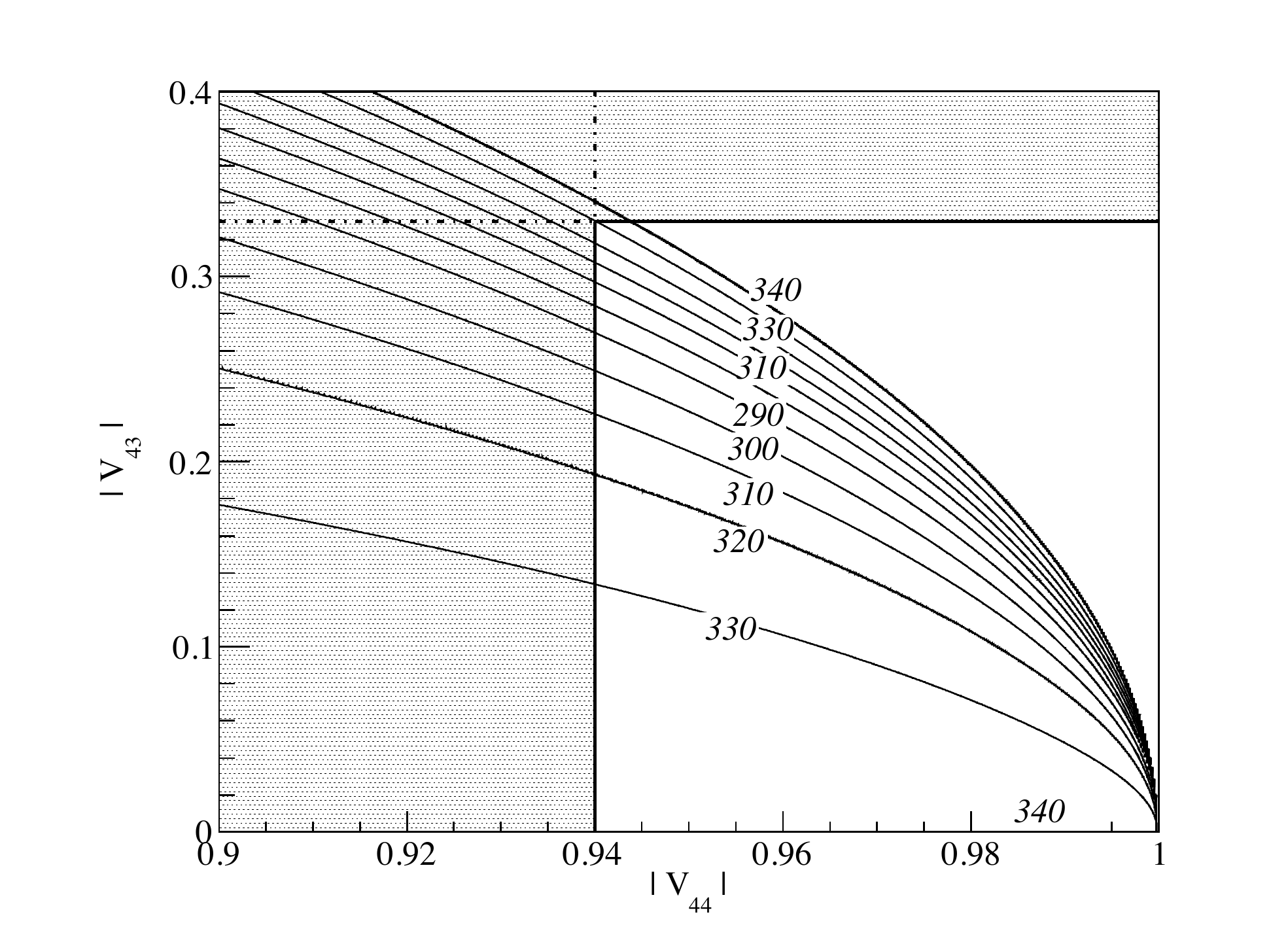}
\includegraphics[width=0.9\linewidth]{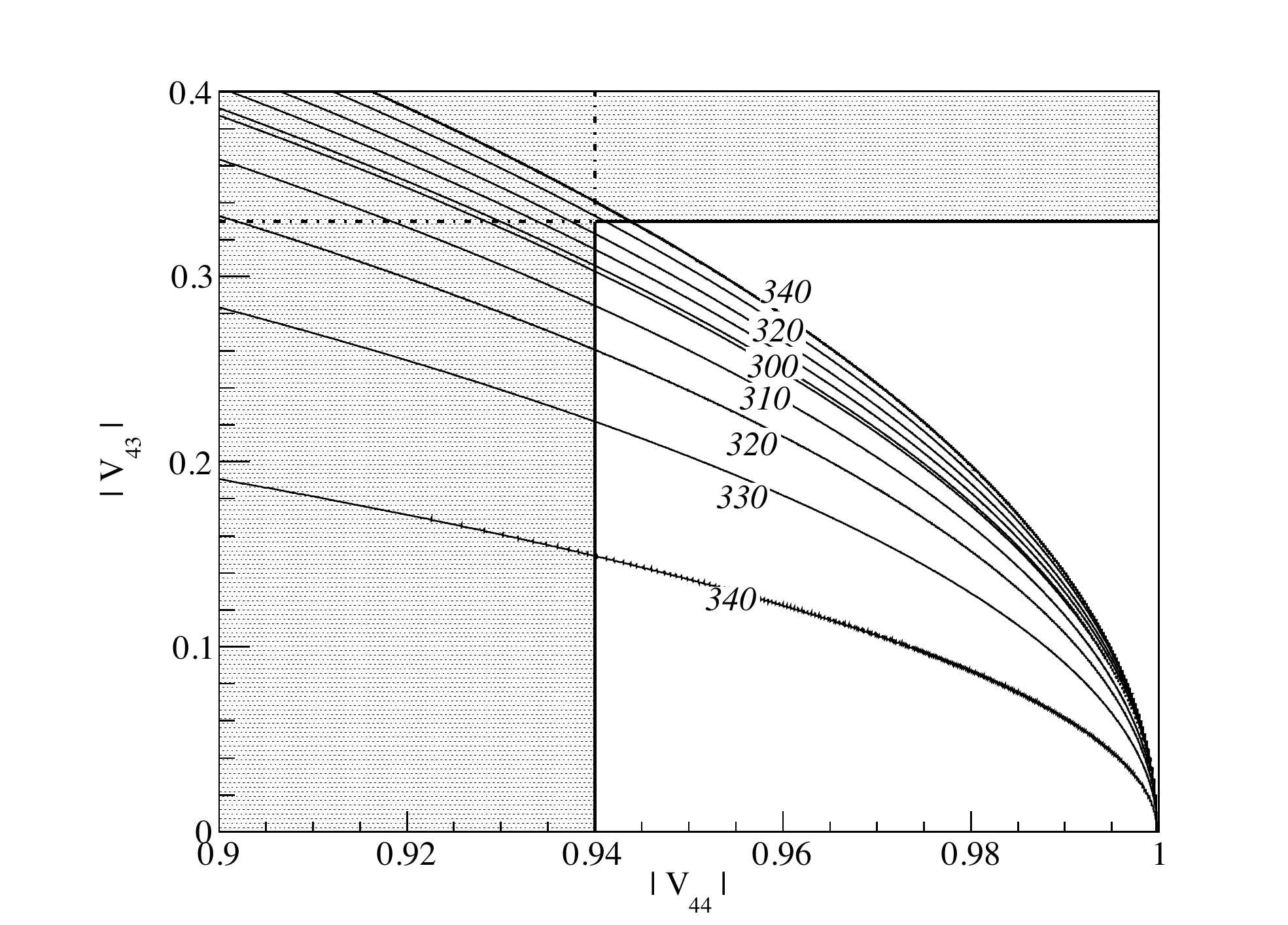}
\caption{ 
  Limits on $b'$ mass from $\ell^{\pm}\ell^{\pm}jb\missET$, \lfivej and
  $\ell+4j$  data, as a function of CKM4 parameters $V_{43}$ and
  $V_{44}$ in the region of interest,
  for mass structures $m_{t'} = m_{b'}+100\ \textrm{GeV}$ (top) and $m_{t'} =
  m_{b'}+50\ \textrm{GeV}$ (bottom).
 }
\label{fig:CKMConts3}
\end{figure}

For the case when the fourth generation $b'$ mixes with the third generation in decay, 
the situation is slightly more complicated. Here we cannot describe the branching
fraction using only the CKM vertex factor $V_{34}$, because the presence of the
kinematically inaccessible (assuming classical splitting) $t'$ nevertheless
affects our expression for the total width when we apply the weak universality
constraint to remove the dependence on mixing with the lighter generations.

\begin{eqnarray}
\{ 1 - \|V_{34}\|^2  - \|V_{44}\|^2 \}\cdot\|\int
^{(b',q)}dp^4(...)\|^2 \nonumber \\ + \|V_{34}\|^2\cdot\|\int^{(b',t)} dp^4(...)\|^2 \nonumber
\end{eqnarray}

In this simplified model, the transformation has only two degrees of freedom associated with 
the phase space integrals. These may be expressed by the ratio of the two factors in each equation. 
This is expected 
since we have reduced the structure of the 
quark masses to two splittings, that between \tprime and \bprime, and between 
\bprime and $t$.

In Fig.~\ref{fig:CKMConst1}, we show the branching ratio as a
function of the two relevant CKM4 parameters,  $V_{43}$ and $V_{44}$
for the mass structure $m_{t'} = m_{b'}+100\ \textrm{GeV}$.
Figure~\ref{fig:CKMConst1B} shows the same for $m_{t'} = m_{b'}+50\
\textrm{GeV}$.  Although strictly speaking the $W$ is virtual in this case,
its behavior in the decay is essentially indistinguishable from the on-shell case.  Finally, we present limits on the fourth generation quarks in the CKM4
space, see Figure~\ref{fig:CKMConts2}.

 The transformation is sensitive to the phase space assumptions, as the amount of available
phase space is mass-dependent. Thus it is important to consider the consistency of
this assumption with the limits set. To demonstrate this, we perform the 
transformation with phase-space assumptions that span the range
of limits set. The degree to which the limit contours in CKM space change
is a measure of the robustness of the transformation. There is very little change as the mass of the $b'$ increases to the the top of the range and beyond. As the $b'$ mass is reduced to 300 GeV and below, some appreciable variation occurs. This variation is shown in 
Fig.~\ref{fig:CKMConts2_alt1} %for the 100 GeV splitting and in Fig.~\ref{fig:CKMConts2_alt2} for the 50 GeV splitting
. Inspection of the figures shows the transformation to be robust over the interval of limits set by these
data.

The region favored by other experimental data is $V_{43}<0.3$ and
$V_{44}>0.94$. Figure~\ref{fig:CKMConts3} shows the limits in this region of interest.

\section{Conclusions}

We reiterate the conclusions of our earlier Letter\cite{lim_prl}:  we find that the CDF data imply limits 
on $m_{\bprime}$ and $m_{\tprime}$ of $290$ GeV and greater over the full range of mixing scenarios, 
for two characteristic choices of the \tprime - \bprime mass splitting: 
$m_{\tprime} > m_{\bprime}$ and $m_{\bprime} > m_{\tprime}$.
The inclusion of a \tprime strengthens the previously obtained 
\bprime mass limit from the \lljbme sample; in
the $m_{\tprime} > m_{\bprime}$ case, by
up to 10\% when $0 < m_{\tprime} - m_{\bprime} < M_W$.
Because of its range of sensitivity, 
the addition of the \lfivej sample, while increasing the mass limits in some neighborhoods
of parameter space, does not significantly improve the overall results of this analysis.

The transformation of the mass limits to CKM space is found to be fairly robust despite the variation
in the phase-space description. Based on electroweak CKM limits\cite{ckm4}, we can identify a region of
branching-fraction parameter space that is most interesting, 
but this does not further constrain the range of
mass limits set by the available data.

\section{Acknowledgements}

We thank Tim Tait, Shaouly Bar-Shalom and Alexander Lenz for useful
conversations.  The authors are supported by grants from the
Department of Energy Office of Science and by the Alfred P. Sloan Foundation.

\end{document}